\documentstyle[12pt]{article}
\def\be{\begin {equation}}
\def\ee{\end {equation}}
\def\bea{\begin{eqnarray}}
\def\eea{\end{eqnarray}}

\def\frac#1#2{{\displaystyle#1\over\displaystyle#2}}

\begin{document}
\vspace{5cm}
\thispagestyle{empty}
\begin{center}

{\LARGE \bf Stochastic Soliton Lattices} 

\vspace{0.5cm}
 
 by

\vspace{0.5cm}

{\Large Gennady A. El ${^1}$}  and  
 {\Large Alexander L. Krylov ${^2}$}

 \vspace{0.5cm}
 Report at the International Conference `Solitons, Geometry
and Topology: on the Crossroads' in honour of the 60-th birthday
of S.P. Novikov,

 Moscow, 1998.

\footnotetext[1] {Institue of Terrestrial Magnetism, Ionosphere
and Radio Wave Propagation, RAS, 142092 Troitsk, Moscow Region, Russia.
e-mail: el@center.izmiran.troitsk.ru}
\footnotetext[2]{O.Yu.Shmidt Institute of Earth Physics, RAS, 123812 Moscow, 
Russia.  e-mail: krylov@td.lpi.ac.ru}
\end{center}

\vspace{1cm}

{\bf Abstract}

We introduce a new concept , {\it Stochastic Soliton Lattice}, as a random
process generated by a finite-gap potential of the Shroedinger operator.
We study the basic properties of this stochastic process 
and consider its KdV evolution  

\section{Introduction}

The finite-gap solutions of completely integrable systems are almost
periodic functions [N], [L], and, therefore, posess their natural stochastic
structure determined  by intrinsic shift group (compact group with the 
Haar measure) [PF]. Ergodic properties of finite-gap potentials
were first used by Flaschka, Forest, and McLaughlin [FFM] in constructing the
theory of multiphase averaging.

The aim of the present work is the consistent introduction of the
basic notions of the finite-gap stochastics, namely, we introduce a
concept of a {\it Stochastic Soliton Lattice} (SSL) and study its 
basic properties. The stochastic structure of  $g$-gap soliton 
lattice (SL)(this term has been introduced by Dubrovin and Novikov in
[DN] for  finite-gap potentials) is the uniform distribution
on $g$-dimensional phase torus (the real section of the correspondent
Jacobian). We also generalize the important notion of the slowly modulated
SL to the stochsatic case. The slow modulations of SSL are described by the
hydrodynamic-type Whitham equations [W], [FFM], [LL], [DN] as well as in the 
case of usual SL.

Following Johnson and Moser [JM] we define the rotation number for the 
SSL. It is well known that the rotation number coincides with the
`density of states' per unit length $\nu (\lambda)$ which is one of the most
important quantities in physics of
disordered systems [LGP].
For the SL the the rotation number is equal to the quasimomentum
[D1] that leads to a considerable effectivization of some general results
of the theory of almost periodic functions.

Passage to the stochastic description is especially natural when
one considers the wave systems with a big number of degrees of freedom.

The case 
of the SSL with $g>>1$ and $\nu <<1$ (weakly interacting rare solitons
Poissonically distributed on the line) corresponds to the 
`soliton turbulence', studied by Zakharov in [Z] invoking the
kinetic description. The similar object is studied by a different method
in the paper of Gurevich, Zybin, and El [GZE] presented in these 
Proceedings.

We note also that the introduced uniform phase distribution
agrees with the Whitham approach to the description of nonlinear
dispersive waves [W], [DN] where the principal term of the 
WKB - decomposition ($g$-gap SL) is determined up to an
arbitrary phase shift.

Calculation of various averaged values on the realization space of 
SSL is an important applied aspect of the proposed theory. 
We present explicit formulas for the one-point probability
function and  effective expressions for two first moments
in the $g$-gap SSL.
The challenge is to find the correlation function (the second mixed
moment).      

\section{Basic Preliminaries}

The soliton lattice (SL) is a function ($g$-gap potential
of the Schroedinger equation)

\be \label{sl}
u_g ( x ; {\bf r})= C ({\bf r}) - 2\partial _{xx}^2 
\ln \Theta [{\bf z (x)}| B({\bf r})] \, ,
\ee

$$
x \in {\bf R}, \qquad {\bf r}=\{r_1, \dots , r_{2g+1}\}, \qquad 
r_1<r_2< \dots < r_{2g}< r_{2g+1} \, . 
$$

Here
\be \label{theta}
\Theta [{\bf z} | B({\bf r})] = \sum \limits_{{\bf m}} \exp \{
\pi i [2({\bf m},{\bf z}) + ({\bf m},B {\bf m})] \}\, ,
\ee
$$
{\bf m}= \{m_1, \dots, m_g \} \in {\bf Z}^g \, , \qquad {\bf z} 
\in {\bf C}^m
$$
is the Jacobi theta-function of the hyperelliptic Riemann surface of 
genus $g$

$$
\mu ^2 = \prod \limits _{j=1}^{2g+1}(\lambda - r_j) \equiv R_{2g+1}
({\bf r},\lambda).
$$
with cuts along the bands.

The Riemann matrix $B({\bf r})$ and the constant $C({\bf r})$ are expressed in 
terms of the basis holomorphic differentials $\psi_j$ :
$$
B_{ij}= \oint \limits _{\beta_j}\psi_i \, , \ \ 
C({\bf r})=\sum_{j=1}^{g} r_j 
-2\sum_{j=1}^{g}\oint \limits_{\alpha _{j}}\lambda \psi_j \, , 
$$
Here
\be \label{psi} 
\psi_j = \sum \limits _{k=0}^{g-1}a_{jk} \frac{\lambda^{k}}{\sqrt { R _{2g+1}
({\bf r}, \lambda)}}d\lambda \, ,
\ee
and dependence $a_{jk}({\bf r})$ is given by the normalization
\be \label{norm}
\oint \limits_{\alpha _k}\psi _{j} = \delta _{jk}\, ,
\ee
where the $\alpha$-cycles surround the bands clockwise, and the $\beta$-
cycles are canonically conjugated to $\alpha$'s.

The imaginary phases ${\bf z}(x)$ are given by the formula

\be \label{z}
{\bf z}=-2i {\bf a}_{g-1}x + {\bf d}\, ,
\ee
where {\bf d} is the initial imaginary phase vector.

Using the substitution
\be \label{y}
{\bf z}=\frac{1}{2\pi}B{\bf y}
\ee
we rewrite (~\ref{sl}) in the form with real phases ${\bf y}$ [FFM]

\be \label{uy}
u_g(x|{\bf r})= u_g (y_1(x), \dots, y_g(x)| {\bf r})\, ,
\ee

\be \label{yj}
y_j(x)=k_jx+f_j \, , \qquad f_j (mod 2\pi) \, ,
\ee

\be \label{k}
{\bf k} = -2\pi iB^{-1}{\bf a}_{g-1}\, .
\ee

Here ${\bf k}= {\bf k}({\bf r})= \{k_1, \dots, k_g \}$
 is the wave number vector, and
${\bf f}= \{f_1, \dots, f_g)\} $ 
is the initial (angle) phase
vector, $-\pi <f_j \leq \pi$.

Note that in (\ref{uy})
\be \label {uq}
u_g(y_1, \dots, y_j+2\pi, \dots, y_g |{\bf r})=
u_g(y_1, \dots, y_j, \dots, y_g |{\bf r})\, ,
\ee
that is $u_g(x|{\bf r})$ is $g$-quasiperiodic on $x$ (from here and on
we consider only incommensurate $k_j, j = 1, \dots, g$. As for any
almost periodic function we have for $u_g(x|{\bf r})$ the Fourier
representation
\be \label{F}
u_g(x|{\bf r})= \sum \limits_{j} c_j e^{i(l_jx +h_j)}\, ,
\ee
where $c_j\, , l_j\, , h_j$ are real, $-\pi<h_j\le \pi$.
(Really, $l_j \in {\bf M} $, where ${\bf M}$ is a frequency modulo [JM] 
and $h_j$ are the integer linear  combinations of $f_j (mod 2\pi)$).

The KdV evolution of (\ref{uy}) is isospectral and is described
by the linear motion of the phases on the Jacobian:
\be \label{uyt} 
u_g(x,t|{\bf r})=u_g(y_1(x,t), \dots, y_g(x,t)|{\bf r})\, ,
\ee

\be \label{yt}
y_j(x,t)=k_jx+ \omega_j t+f_j \, ,
\ee
where 
\be \label{omega}
{\bf \omega}= {\bf \omega}({\bf r}) = \{\omega_1, \dots, 
\omega_g \} = -4\pi i B^{-1} ({\bf a}_{g-1}\sum \limits_{j=1}^{2g+1}r_j
+ 2{\bf a}_{g-2})
\ee
 is the frequency vector.

We call $u_g(x|{\bf r})$ in (\ref{uy}), (\ref{yj}) the modulated SL
if ${\bf r}= {\bf r}(\epsilon x)\, , \epsilon
\ll 1$ (the possibility of
$g = g(\epsilon x)$ is included).
The initial value problem for the KdV equation

\begin{equation}\label{KdV}
u_t - 6 uu_x + u_{xxx} = 0 \, ,
\end{equation}

\be \label{u0}
u(x,0) = u_g(x|{\bf r}(\epsilon x))
\ee
has the solution 
\be \label{umod}
u(x,t) = u_g(x,t|{\bf r}(\epsilon x , \epsilon t))\, ,
\ee
where $u_g(x,t|{\bf r})$ is given by (\ref{uyt}),(\ref {yt})
and the evolution of modulation parameters ${\bf r}(X, T)$
where $(X,T)= ( \epsilon x , \epsilon t)$ is given by the 
Whitham equations [W],[FFM],[LL],[DN]

\begin{equation}\label{Wh} 
\partial _T r_j + V_j({\bf r}) \partial_ X r_j = 0 \, ,
\qquad\qquad j=1,\dots, 2g+1 \, ,
\end{equation} 

\be \label{r0}
r_j(X,0) = r_j(X)\, .
\ee
Here $g=g(X,T)\, , \ g(X,0) =g(X)$ and the Whitham equations (\ref {Wh}),(\ref{r0})
must be supplemented with the additional information about evolution of $g$
[LL], [V1], [DVZ], [D1].

\section{Stochastic Soliton Lattices: Definitions and Basic Properties}

It is well known that any almost periodic function generates the
stochastic stationary process (see [JM], [PF]).

{\bf Definition 1.} The stochastic process generated by SL $u_g(x|{\bf r})$
we call {\it Stochastic Soliton Lattice} (SSL) and denote as 
$\nu _g (x|{\bf r})$.

The general construction of $\nu _g (x|{\bf r})$
admits a very simple and clear description. The realization set 
$\Omega$ of it consists of functions (\ref{uy}), (\ref{yj}) where
${\bf f}\in T^g$ ; $T^g$ is $g$-dimensional torus $(-\pi\, ,\pi]^g$.
The probability measure $d\mu$ is the uniform (Lebesque) measure 
on the torus (on Borel sets on it). It corresponds to the description
of $\nu _g(x|{\bf r})$ following from(\ref{uy}),(\ref{yj}):
\be \label{ssl}
\nu _g (x|{\bf r}) = u_g (\dots k_j x+\phi _j \dots|{\bf r}) \, \
\ee
where $\phi_1, \dots , \phi_g$ are independent random values
uniformly distributed on $(-\pi\, , \pi]$, that is ${\bf \phi}=\{\phi_1,
\dots, \phi_g \}$ is uniformly distributed on $T^g$. As $k_j$ are
incommensurate then $\nu _g (x|{\bf r})$ is an ergodic process [CSF].

As $\nu _g (x|{\bf r})$ is the stationary process then it has the 
Stone - Kolmogorov spectral decomposition [I]; due to the ergodicity
this decomposition has the form (cf. (\ref {F}))
\be \label{SK}
\nu _g (x|{\bf r})= \sum \limits_{j} c_j e^{i(l_jx +\theta_j)}\, ,
\ee
where $\theta _j$ are the uniformly distributed on $(-\pi \, , \pi]$
{\it noncorrelated} random values [PR].

The well known formula (Bochner - Khintchin) [I] gives us the
covariance function $K(h)$ of the stationary process $\nu _g (x|{\bf r})$:
\be \label{cov}
K(h)\equiv \langle \hat{\nu }_g (x|{\bf r})\cdot 
\hat{\nu} _g (x|{\bf r})\rangle =
\sum _j |c_j|^2 e^{i\lambda_j x}\, .
\ee
Here $\hat{\xi}(x) \equiv \xi(x) - \langle \xi \rangle$ is the
centered process.

{\bf Theorem 1.}  Consider the KdV equation (\ref{KdV}) as the
equation describing the evolution in the phase space of stationary 
processes. Let the initial data has the form of the SSL:
\be \label{nu0}
u(x,0)= \nu _g (x|{\bf r}) \, .
\ee
Then the solution of the KdV is
\be \label{nut}
u(x,t)= \nu _g (x|{\bf r})= u(x,0)\, ,
\ee
that is $u(x,0) = \nu _g (x|{\bf r})$ is a stationary point.

{\bf Proof . }The evolution of realizations (\ref{uy}) ,(\ref{yj})
is described by (\ref{uyt}) ,(\ref{yt}). For any moment $t$ one can
introduce the new `initial phase' $f^*_j=\omega _j t +f_j$ which is
also uniformly distributed on the torus $T^g$. Therefore, the KdV
evolution changes neither realization set $\Omega$ nor the probability
measure.

\noindent {\bf Remark.} In the connection with the Theorem 1 what can one say
about the closure (in some topology) of the set $\{\nu _g (x|{\bf r}),
{\bf r} \in {\bf R}^g, g \in {\bf Z}\}$?

{\bf Definition 2.} We call $\nu _g (x|{\bf r})$ from
(\ref{ssl}) a {\it modulated SSL}
if ${\bf r}={\bf r}(\epsilon x)$.

The modulated SSL (in opposite to the nonmodulated one) does evolve
according to the KdV.

{\bf Theorem 2.} Consider the KdV intial value problem in the
sense of the Theorem 1 with the same form of the initial data
(\ref{nu0}) but with ${\bf r}={\bf r}(\epsilon x)$.
Then the solution of it is  
\be \label{num}
u(x,t) = \nu _g (x|{\bf r}(\epsilon x, \epsilon t)) \, ,
\ee
where ${\bf r}(X,T)$ is described by (\ref{Wh}),(\ref{r0}).

\noindent {\bf Remark .} In a more general case the modulation function
${\bf r}(X)$ can be random itself (the simplest example is 
almost periodic ${\bf r}(X)$). 

\section{Some Mean Values (Moments) for SSL}

If $Q(f)$ is an arbitrary smooth function then the mean value $Q(\xi)$
of the stochastic process $\xi$ is

$$
\langle Q(\xi)\rangle = \int \limits_{\Omega} d\mu Q(f) \, , \qquad
f \in \Omega \, . 
$$ 

(We recall that $\xi(x) \equiv \{\Omega = \{f(x)\}\, , B, \mu \}$,
and $B$ is some $\sigma$-algebra of measurable Borel sets 
of $\Omega$).

For $\nu _g (x|{\bf r})$ we have [FFM]

\be \label{ergod}
\langle Q(\nu _g (x|{\bf r})\rangle = \frac{1}{(2\pi)^g}
\int \limits ^{\pi}_{-\pi}\dots \int \limits ^{\pi}_{-\pi}
d\phi_1 \dots d\phi_g Q(\nu _g (x|{\bf r}))=
\lim \limits_{L \to \infty }\frac{1}{L} \int \limits ^{L} _{0}
Q(\nu _g (x|{\bf r})) \, .
\ee
 
 Direct calculation using (\ref{sl}),(\ref{theta}) gives
 surprisingly simple formulas for the mean value [V3] and the mean
 square of
 \be \label{nutilda}
 \tilde{\nu} _g (x|{\bf r})= \nu _g (x|{\bf r}) - C({\bf r})\, .
 \ee
 Namely,
 \be \label{mean}
 \langle \tilde{\nu }_g (x|{\bf r})\rangle = \frac{1}{4\pi i}({\bf k},B{\bf k})\,
 , \qquad
 \langle \tilde{\nu} _g ^2(x|{\bf r}) \rangle = - 
 \frac{1}{4\pi i}({\bf \omega},B{\bf k})\,  ,
 \ee 
 and ${\bf k}({\bf r}), {\bf \omega}({\bf r})$ are given by
 (\ref{k}) , (\ref{omega}). 
 It should be noted that expressions (\ref{mean}) are obtained for the
 particular choice of the canonical basis of cycles (see (\ref{norm})).
 However, namely this normalizatiton is preferrable for many considerations
 of finite-gap solutions (see for example [FFM], [V2]). 
 
 We still do not know the expression for more important value: 
 the second mixed moment $\langle \tilde{\nu} _g (x|{\bf r})\tilde{\nu} _g
 (x+h|{\bf r}) \rangle$.
  
  \section {One - Point Generation Function for the Moments }
  
  The function
  \be \label{exp}
  f(\lambda)=\langle e^{i\lambda u}\rangle 
  \ee
  gives the moments $\langle u\rangle, \langle u^2\rangle, \dots$
  as the coefficients in expansion of $f(\lambda)$ in powers
  of $\lambda$. Obviously, the one-point probability function
  is connected with $f(\lambda)$ by a Fourier transform:
  \be \label{prob}
  w(u)=\frac{1}{2\pi} \int \limits _{-\infty}^{\infty}
  f(\lambda) e^{-i\lambda u} d \lambda 
  \ee
  
 To calculate $f^g(\lambda)$ for the SSL we make use of the trace formula
 for the SL $u=u_g({\bf y})$(\ref{uy}) [Lev], [FFM]
 \be \label{trace}
 u_g=r_{2g+1} - 2\sum \limits_{j=1}^{g}(\mu_j +\eta_j^2) \, ,
 \ee
 where
 \be \label {cband}
\eta_j ^2 =-\frac{r_{2j-1} + r_{2j}}{2}
\ee 
 is the central point of the band and the functions $\mu _j$ obey
 the Dubrovin's ODE's [D2] and each $\mu_j$ lives inside the $j$ -th gap.
 Then the ensemble integral (\ref{ergod}) can be represented as
 $$
 \langle e^{iu\lambda} \rangle =\frac{1}{(2\pi)^g}
 \int \limits ^{\pi}_{-\pi}\dots \int \limits ^{\pi}_{-\pi}
 e^{i\lambda u_g({\bf y})} dy_1 \dots dy_g = 
 $$
 \be \label{int}
 =\frac{1}{(\pi)^g}e^{i\lambda r_{2g+1}}
 \int \limits ^{r_3}_{r_2}\dots \int \limits ^{r_{2g+1}}_{r_{2g}}
 e^{-2i\lambda \sum \limits_{j=1}^{g}(\mu_j+{\eta_j}^2)}
 \frac{\partial y}{\partial \mu}d\mu_1 \dots d\mu_g \, .
 \ee
 The Jacobian $\frac{\partial y}{\partial \mu}$ appears in [FFM]
 and has the form
 \be \label{Jac}
 \frac{1}{\pi ^g} \frac{\partial y}{\partial \mu}=
 \frac{\det \frac{\mu ^{j-1}_l}{\sqrt {R(\mu_l)}}}{\det[
 \int \limits^{r_{2l+1}} _{r_{2l}}\frac{\mu^{k-1}}{\sqrt{R(\mu)}}d\mu]} \, .
 \ee
 
 Substitution of (\ref{Jac})into(\ref{int}) allows to represent
 the multiple integral in the form of the product
 \be \label{f}
 f_g(\lambda)= e^{i\lambda r_{2g+1}}
 \frac{\det [\int \limits^{r_{2l+1}}_{r_{2l}}
 \frac{e^{-2i\lambda (\mu + \eta_l^2)}\mu^{j-1}}{\sqrt{R(\mu)}}d\mu]}
 {\det [\int \limits^{r_{2l+1}}_{r_{2l}}
 \frac{\mu^{j-1}}{\sqrt{R(\mu)}}d\mu]} \, .
 \ee
 
 Taking the Fourier transform of (\ref{f}) we get the probabilty density
 $w_g(u)$(\ref{prob}). (As a matter of fact, $\int \limits_{-\infty}
 ^{\infty} w_g(u)du = 1$ as $f_g(0)=1$).
 
 The next step is to calculate the two-point generating function
 $f_g(\lambda_1, \lambda_2; h) $
 \be \label{2point}
 f_g(\lambda_1, \lambda_2; h) = \langle e^{-i\lambda_1 u(x)}
 e^{-i\lambda_2 u(x+h)}\rangle \, ,
\ee
 which gives  the covariance function (\ref{cov}) as the 
 term before $\lambda_1 \lambda_2$ in the decomposition
 of (\ref{2point}).

\section{Density of States (Rotation Number) for SSL}

Consider the Schroedinger equation with almost periodic
potential $q(x)$

\be \label{shr}
L\phi=(-\partial ^2 _{xx}+ q(x))\phi=\lambda \phi \, ,
\qquad  x \in {\bf R} \, .
\ee

The potential $q(x)$ has a very important characteristics, 
 {\it the rotation number}, which is defined for real $\lambda$
as [JM] 
\be \label{rot}
 \alpha(\lambda)=\lim \limits _{x \to \infty} \frac{1}{x}
 \arg (\phi \prime (x,\lambda) - i\phi(x,\lambda))\, .
 \ee 

We will use also the {\it density of states} $\rho (\lambda)$
which is connected with the rotation number by
simple relation 
\be \label{dens}
\rho (\lambda)= \frac{1}{\pi} \alpha(\lambda)
\ee
and the {\it integral density of states} [LGP], [PF],
\be \label{N}
N(\lambda) \equiv \int \limits _{-\infty}^{\infty} \rho(\lambda)d\lambda\, .
\ee
If $q(x)$ is a finite-gap potential , $q(x)=u_g(x|{\bf r})$,
then [D1]
\be \label{ap}
\alpha(\lambda)=\frac{dp(\lambda)}{d\lambda}\, ,
\ee

where $dp(\lambda)$ is the quasimomentum [FFM], [DN] normalized
by
\be \label{pnorm}
\oint \limits _{\beta _j}dp(\lambda)=0 \, , \qquad j=1,\dots , g \, .
\ee

As the value of $\alpha(\lambda)$ is the same for any realization
from the set $\Omega$ from Def.1 then it is the spectral 
characteristics of the whole SSL $\nu(x|{\bf r})$. Due to the Theorem 1
the rotation number does not change under the KdV evolution.

We introduce also the {\it full integral density } of states in the SSL
$\nu$ as
\be \label{plot}
\nu \equiv  N(\infty)=\frac{1}{\pi}\int \limits_{-\infty}^{\infty}
dp(\lambda)=\frac{1}{2\pi}\sum _{j=1}^{g}\oint \limits _{\alpha _j}
dp(\lambda)=\frac{1}{2\pi}\sum _{j=1}^g k_j \, .
\ee
where $k_j$ are the components of the wave number vector (\ref{k}).
Here we have used the known relationship [FFM],[DN] (recall that the
$\alpha$ - cycles surround the bands in our normalization (see Sec.2))
\be \label{kj}
 \oint  \limits _{\alpha_j} dp(\lambda)=k_j \, .
\ee
 One can see that the full integral density of states has here the natural
 meaning of the mean number of waves per unit length.

\noindent {\bf Remark.}

\noindent As the SL (\ref{uy}), (\ref{yj})is the quasiperiodic
function both in $x$ and $t$ we can deduce that an analogous result
is valid for the {\it full temporal density of states} as well, namely
\be \label{tplot}
\nu ^t = \frac{1}{2\pi}\sum _{j=1}^g \omega _j \, ,
\ee
where $\omega _j$ are the frequences (\ref {omega}),
that implies that the {\it temporal rotation number} in the SSL is
\be
\alpha ^t (\lambda)= \frac{dq(\lambda)}{d\lambda}\, ,
\ee 
where $dq(\lambda)$ is the quasienergy [FFM],[DN] ( $\oint  \limits _{\alpha_j}
dq(\lambda)=\omega_j$).
\section {Distribution of Random Initial Phases in SSL.\\ 
Poisson and Normal Limits} 

We introduce the {\it linear phases} 
\be \label {l}
l_j \equiv \frac {\phi_j}{k_j}\, ,
\ee
where $l_j \, \ (j=1, \dots, g)$ are independent random values uniformly
distributed on $(-\frac{\pi}{k_j}, \frac{\pi}{k_j}]$ each.

Introduce also the random value $\xi_j = \chi _{(0,1)}(l_j)$ 
which is the number of hitting 
of $l_j$ into the fixed interval $(0,1) \subset {\bf R}$ (we suppose
that $\pi /{k_j}\geq 1$). The variable $\xi _j$ takes two values: $1$ and $0$
with the probabilities $p_j(1)= 1/{2\pi}$ and
$p_j(0)=q_j=1-p_j=1-{k_j}/{2\pi}$ . 
The generating function $\varphi _j(z)$ of $\xi_j$ is

\be \label{phij}
\varphi _j(z)=(1-p_j) + zp_j
\ee

The sum $\xi^{(g)}\equiv \sum \limits _{j=1}^g \xi_j$  is the number of 
hitting of all linear phases into $(0,1)$. As $\xi_j$ are independent,
the generating function
for $\xi^{(g)}$ has the form 
\be \label{phig}
\varphi^{(g)}(z)= \prod _{j=1}^{g}\phi _j(z)= \prod _{j=1}^{g}
(1+(z-1)p_j)=\prod _{j=1}^{g}(1+\frac{(z-1)k_j}{2\pi}) \, . 
\ee

Consider now the important particular case $g \gg 1$
There are two subcases to consider.

\subsection{Finite full integral density of states}

Suppose the following scaling
\be \label{scale}
k_j=O(g^{-1})\, .
\ee
Then taking the logarithm of (\ref{phig}) we obtain
$$
\ln \varphi^{(g)}(z)= (z-1)\frac{1}{2\pi}\sum _{j=1}^{g}k_j + O(g^{-1}) \, .
$$
Therefore
\be \label{phii}
\varphi^{(\infty)}(z)=\exp\{(z-1)\nu \}\, .
\ee
It is well known that the right-hand
part of (\ref{phii}) is the generating function for 
the {\it Poisson distribution} with the mean value $\nu$.
That means that  $\xi^{(\infty)}$
is a Poissonic random value with the full integral density
of states as a mean value.

\noindent{\bf Remark 1.}

\noindent The scaling (\ref{scale}) arises in particular 
in the semiclassical asymptotics for  periodic KdV potentials. 
potentials [V1][WK].

\noindent {\bf Remark 2.}

\noindent If $\nu \ll 1$ then we arrive at the case of the separate
oscillations asymptotically close to solitons Poissonically distributed 
on the line. The linear phases $l_j$ in this case give the asymptotic 
positions of the "solitons". This case corresponds to Zakharov's `soliton 
turbulence' [Z].

\subsection{Large full integral density of states}

Suppose now that the following conditions fulfil

\noindent 1. $\sum _{j=1}^{g} k_j =\nu \gg 1$,

\noindent 2. $k_j = o(1)$.

\noindent Then the value $\xi^{(g)}$ is distributed asymptotically 
as $g \to \infty$ by a normal distribution. 
This follows from the Lyapunov Central Limit Theorem in Berry - Essen
formulation [F]. We present here the corresponding estimates for the 
distribution  function $F_g(x)$ for the normed sums 

$$
\frac{\sum_{k=1}^{g} \xi_k - \sum_{k=1}^{g} p_k}
{(\sum_{k=1}^g p_k q_k)^{1/2}} \, .
$$

The estimate is
\be \label{}
| F_g(x)-\Phi (x)|= O(g^{-1/2})\, ,
\ee
where
$$
\Phi(x)=\frac{1}{2\pi}\int \limits _{-\infty}^{x} e^{-y^2/2} dy \, .
$$ 

\vspace{1cm}
{\bf Aknowledgments.}

We thank Stephanos Venakides for useful discussions.

The research described in this paper was made possible  by the
financial support of the 
Civilian Research and Development Foundation for the Former
Soviet Union and USA (CRDF) under the grant $\#$ RM1 - 145.

We also thank the Russian Foundation for Basic Research (RFBR) for the
partial finansial support under the grant $\#$ 96 - 01 - 01453. 

\vspace{1cm}

\begin{center}{\bf REFERENCES}
\end {center}
\vspace{0.5cm}

{\bf [CSF]} I.P.Cornfeld, Ya.G.Sinai, S.V.Fomin, {\it Ergodic theory }(1980) 
Nauka, Moscow. 

{\bf [DVZ]}P.Deift, S.Venakides, and X.Zhou, International Math. Research. 
Journ., No 4 (1997) 285.

{\bf [D1]} B.A.Dubrovin, Amer.Math.Soc.Transl (2) {\bf 179} (1997) 35.

{\bf [D2]} B.A.Dubrovin,  Russian Math. Surveys  {\bf 36} (1981) 215.

{\bf [DN]} B.A.Dubrovin and S.P.Novikov, Russian  Math. Surveys {\bf 44} (1989)
35 .

{\bf [F]} W.Feller {\it Introduction to the Probability Theory}

{\bf [FFM]}  H. Flaschka, G. Forest, D.W. McLaughlin, Comm. Pure Appl.
Math. {\bf 33} (1979)739 . 

{\bf [GZE]} A.V.Gurevich, K.P.Zybin, and G.A.El,
 {\it Development of stochastic oscillations
in one-dimensional dynamical system described by the
Korteweg -- de Vries equation}, these Proceedings.  

{\bf [I]} Ito, {\it Stochastic Processes}, (1970) Mir, Moscow.

{\bf [JM]} R.Johnson and J.Moser, Comm.Mah.Phys., {\bf 84} (1982) 403.

{\bf [L]} P.D. Lax, Comm.Pure Appl. Math. {\bf 26} (1975) 141.

{\bf [Lev]} B.M.Levitan , {\it Inverse Sturm-Liouville problems},(1984), 
Nauka, Moscow.

{\bf [LL]} P.D.Lax and C.D.Levermore, Comm. Pure Appl. Math. {\bf 36} (1983) 
253,571,809.

{\bf [LGP]} I.M.Lifshitz, S.A.Gredeskul, A.L.Pastur, {\it Introduction to the 
disordered systems theory}, (1982) Nauka, Moscow .

{\bf [N]} S.P.Novikov, Func. Anal.Pril., {\bf 8} (1974) 54.

{\bf [PF]} L.A.Pastur, A.L.Figotin, {\it Random and almost periodic
self-adjoint operators}, (1991) Nauka, Moscow.

{\bf [PR]} Yu.V. Prokhorov, Yu.A.Rozanov, {\it Probability Theory},
Nauka, Moscow, 1970.

{\bf [V1]} S. Venakides, AMS Transaction {\bf 301} (1987)  189.

{\bf [V2]} S. Venakides, Comm.Pure Appl.Math. {\bf 42} (1989) 711. 

{\bf [V3]} S. Venakides, private communication.

{\bf [WK]} M.I.Weinstein, J.B.Keller, SIAM Jornal Appl.Math.  {\bf 47},
941 .

{\bf [W]} G.B.Whitham, Proc. Roy. Soc. {\bf A283} (1965) 238.

{\bf [Z]} V.E.Zakharov, Sov.Phys.JETP {\bf 60} (1971) 1012 [In Russian].

\end{document}